\documentclass{webofc}
\usepackage[varg]{txfonts}   

\usepackage{graphicx}
\usepackage{subcaption}

\begin{document}

\title{Decay of the $\Lambda(1405)$ Hyperon to $\Sigma^0\pi^0$ Measured at GlueX}

\author{\firstname{Nilanga} \lastname{Wickramaarachchi}\inst{1}\fnsep\thanks{\email{wickramaarachchi@cua.edu}} \and
        \firstname{Reinhard A.} \lastname{Schumacher}\inst{2}\and
        \firstname{Grzegorz} \lastname{Kalicy }\inst{1} 
        ~for the GlueX Collaboration
}

\institute{The Catholic University of America, Washington, D.C. 20064, USA
\and  Carnegie Mellon University, Pittsburgh, PA 15213, USA }

\abstract{
 Among the light baryons, the $J^\pi = \frac{1}{2}^-$ $\Lambda(1405)$ hyperon is an important special case as it sits just below the $N\bar{K}$ threshold and decays almost exclusively to $\Sigma\pi$. Some long-standing hypotheses for $\Lambda(1405)$ are that it could be a $N\bar{K}$ bound state or a $\Sigma\pi$ continuum resonance. It may also be considered as a simple quark-model resonance, the $P$-wave companion of the $\Lambda(1520)$. In recent years chiral unitary models have suggested that there are two isospin zero poles present in this mass region, and that the ``line shape" of the $\Lambda(1405)$ depends to what extent each of the two poles are stimulated in a given reaction. Below the $N\bar{K}$ threshold, the $\Lambda(1405)$  decays to the three $\Sigma\pi$  charge combinations. The $\Sigma^{0}\pi^{0}$ mode  is purely $I=0$ and it is uncontaminated by complications arising from $I=1$ scattering processes contributing to the reaction mechanism in the $\Sigma^{+}\pi^{-}$ and $\Sigma^{-}\pi^{+}$ decays. It is also not affected from production and decay of the  nearby $\Sigma^{0}(1385)$  hyperon.  The GlueX experiment at Jefferson Lab has been used to study the $\Lambda(1405) \to \Sigma^{0}\pi^{0}$ decay mode. We focus on the preliminary results of $d\sigma/dM_{\Sigma^{0}\pi^{0}}$ and fits to the line shape of the $\Lambda(1405)$ region in the $-(t-t_{min})$ range 0 - 1.5 GeV$^2$ from analyzing the reaction $\gamma p \to K^{+}\Lambda^*$ using the data collected during the first phase of the GlueX program. \\
}

\maketitle
\section{Introduction}
\label{Sec:intro}

The $\Lambda(1405)$ is a hyperon with $J^\pi = \frac{1}{2}^{-}$ and located just below the $N\bar{K}$ threshold. Observation of this state dates back to 1961 where it was produced by $K^{-}p$ interactions
using a 1.15~GeV/$c$ momentum $K^{-}$ beam in bubble chamber experiments~\cite{RefKmp}. Because of its proximity to the  $N\bar{K}$ threshold, it was predicted on the basis of hadronic scattering data to be a lightly bound ``molecule" of these particles~\cite{Dalitz}.  Since then the $\Lambda(1405)$ has been widely discussed to understand it's nature. In the quark model the $\Lambda(1405)$ can be considered as the spin-orbit partner of $\Lambda(1520)$. One of the interesting features of the $\Lambda(1405)$ is its invariant mass spectrum, also called its ``line shape". The line shape in a given production reaction may give clues to the internal structure of the state.  It has been measured by experiments including CLAS~\cite{RefCLAS}, ANKE~\cite{RefANKE} and BGO-OD~\cite{RefBGOOD}, which show distortions from a simple resonance form. The cause for this behavior is not fully understood, but new measurements at GlueX may help clarify the picture.\\ 

Below the $N\bar{K}$ threshold, the $\Lambda(1405)$ decays 100\% into $\Sigma\pi$, with each mode $\Sigma^{+}\pi^{-}$, $\Sigma^{-}\pi^{+}$ and $\Sigma^0\pi^0$ expected to have close to 1/3  branching fraction. Above the $N\bar{K}$ threshold, the $\Lambda(1405)$ is assumed to couple strongly to this channel. Various  chiral unitary models suggest that the $\Lambda(1405)$ is a compound state consisting of two $I=0$ poles;  for a recent discussion, see Ref.~\cite{RefChiral}. The line shape can depend on how each of the two poles are contributing to the reaction. In fact, the most recent PDG has added a $\Lambda(1380)$ as a two-star resonance to account for this two-pole picture of $\Lambda(1405)$~\cite{RefPDG}. Among the decay modes of the  $\Lambda(1405)$,~$\Sigma^{0}\pi^{0}$ is a special case since it is purely $I=0$. This is shown in Eq.~\ref{Eqn:pipSigmam}-\ref{Eqn:pi0Sigma0} where the differential cross section for each mode is written in terms of contributions from different isospin amplitudes~\cite{RefIsospin}.

\begin{equation} \label{Eqn:pipSigmam}
\frac{d\sigma(\pi^{+}\Sigma^-)}{dM_{I}} \propto \frac{1}{3}|T^{(0)}|^2 + \frac{1}{2}|T^{(1)}|^2 + \frac{2}{\sqrt{6}}\text{Re}(T^{(0)}T^{(1)*})
\end{equation}

\begin{equation} \label{Eqn:pimSigmap}
\frac{d\sigma(\pi^{-}\Sigma^+)}{dM_{I}} \propto \frac{1}{3}|T^{(0)}|^2 + \frac{1}{2}|T^{(1)}|^2 - \frac{2}{\sqrt{6}}\text{Re}(T^{(0)}T^{(1)*})
\end{equation}

\begin{equation} \label{Eqn:pi0Sigma0}
\frac{d\sigma(\pi^{0}\Sigma^0)}{dM_{I}} \propto \frac{1}{3}|T^{(0)}|^2
\end{equation}\\

Unlike the two charged modes there is no contamination from the $I=1$  $\Sigma(1385)$ to the $\Sigma^0\pi^0$ mode. So reconstructing $\Lambda(1405)$ in the $\Sigma^0\pi^0$ mode is very useful to study its line shape. Studying the $\Lambda(1405)$ line shape would provide more information on how the $\Sigma\pi$ and $N\bar{K}$ channels contribute to its production.

\section{Experimental procedures}
\label{Sec:Exp}
\subsection{GlueX Detector}
\label{Sec:GlueX}

The GlueX detector is located in Hall D at Jefferson Lab, Newport News, Virginia, USA. It uses an 11.6 GeV electron beam provided by the Continuous Electron Beam Accelerator Facility (CEBAF). The electron beam is incident on a thin diamond radiator and produces bremsstrahlung photons. The scattered electron is measured and used to tag the energy of the beam photon in the energy range 3.0-11.6 GeV. A fraction of the photons which are $\sim$9 GeV are linearly polarized and produced by coherent bremsstrahlung. While this is important for the main goals of the GlueX experiment, particularly the search for hybrid mesons, the present study does not exploit the linearly polarized beam. The tagged photon beam is incident on a 30 cm long liquid hydrogen target within a superconducting solenoid magnet that provides a 2 T magnetic field. The GlueX detector includes forward and cylindrical drift chambers to track charged particles. Forward and cylindrical electromagnetic calorimeters are used to detect neutral showers and to provide triggering information. Precision timing information from the forward time-of-flight system and the barrel calorimeter is used for particle identification. GlueX detector provides nearly 4$\pi$ angular coverage and allows the exclusive reconstruction of many reactions. Complete information on the GlueX detector can be found in Ref.~\cite{RefGlueX}. Phase I of the GlueX experiment ran during 2017-2018. The total luminosity for the photon beam energy range 6.5-11.6 GeV is 423 pb$^{-1}$.

\subsection{Analysis method}
\label{Sec:Analysis}

\begin{figure}[h]
\centering
\includegraphics[width=10cm,clip]{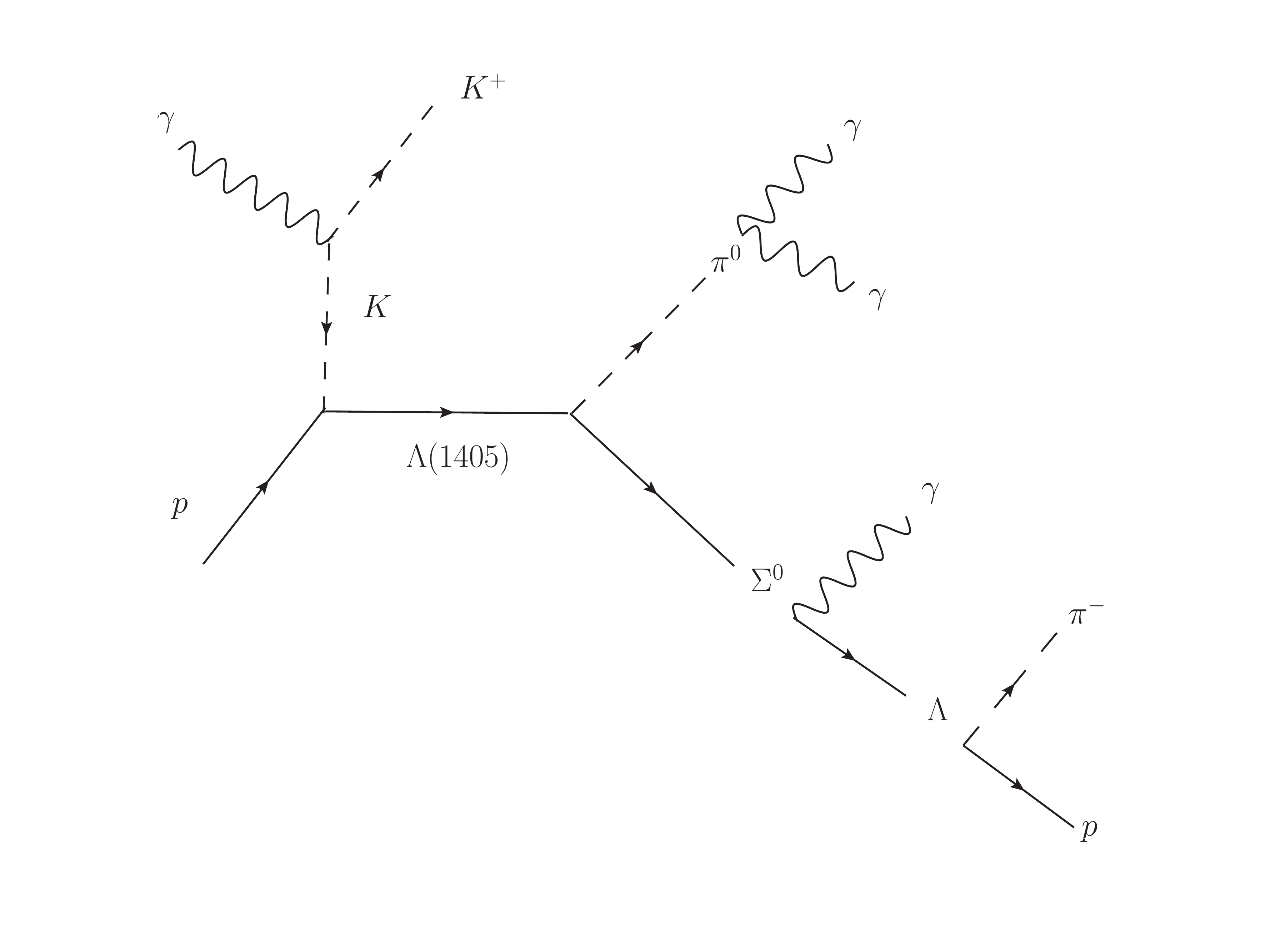}
\caption{The decay chain of the reaction $\gamma p \to K^{+}\Lambda(1405)$.}
\label{fig:reaction}      
\end{figure}

The $\Lambda(1405) \to \Sigma^0\pi^0$ decay was reconstructed using the exclusive reaction $\gamma p \to K^+\Sigma^0\pi^0$. The decay chain is $\Sigma^0 \to \Lambda \gamma$, $\pi^0 \to \gamma \gamma$ and $\Lambda \to \pi^- p$ as shown in Fig.~\ref{fig:reaction}. The final state for this reaction contains two positively charged tracks ($K^+$ and proton), one negatively charged track ($\pi^-$) and three neutral showers (3$\gamma$). A kinematic fit was applied conserving four-momentum of the reaction and with a vertex constraint to allow for displaced vertices of the hyperons. The $\pi^0$ and $\Sigma^0$ masses were constrained during the kinematic fit to improve mass resolution for $\Sigma^0\pi^0$. The full beam energy range $E_{beam} > 6.5$ GeV was used for the analysis in order to apply the measured photon flux to produce cross section $d\sigma/dM_{\Sigma^0\pi^0}$. The Mandelstam $t$ was calculated using four-momenta of the photon beam and the $K^{+}$, $t = (p_{beam}-p_{K^{+}})^2$. Events were selected in the range $ 0 < -(t-t_{min}) < 1.5$~\text{GeV}$^2$ focusing on the $t$-channel production of $\Lambda(1405)$.

\begin{figure}[h]
\centering
\begin{subfigure}[b]{.5\textwidth}
  \centering
  \includegraphics[width=6.7cm,clip]{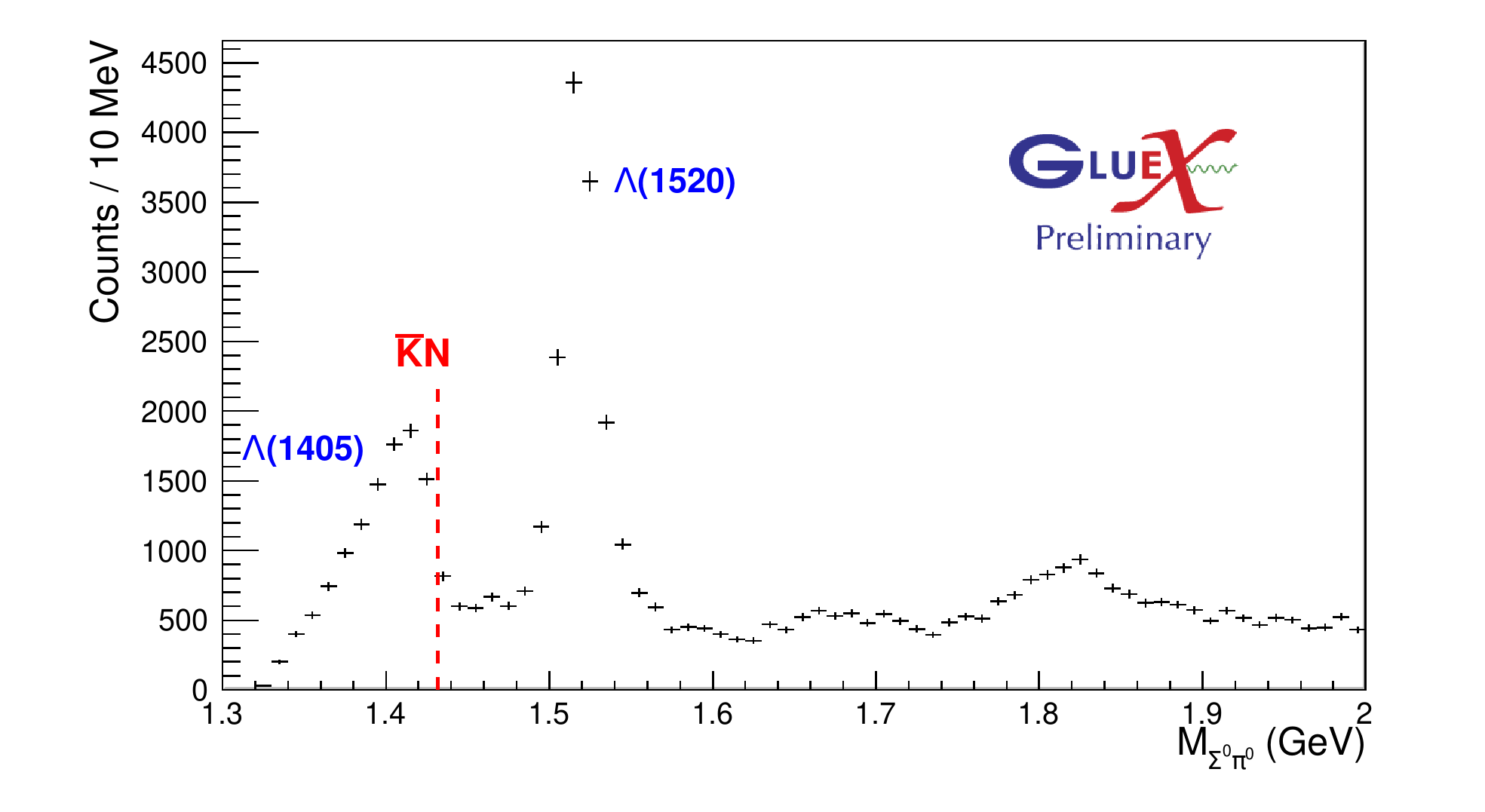}
  \captionsetup{justification=centering}
  \caption{}
  \label{fig:IM}
\end{subfigure}%
\begin{subfigure}[b]{.5\textwidth}
  \centering
  \includegraphics[width=6.7cm,clip]{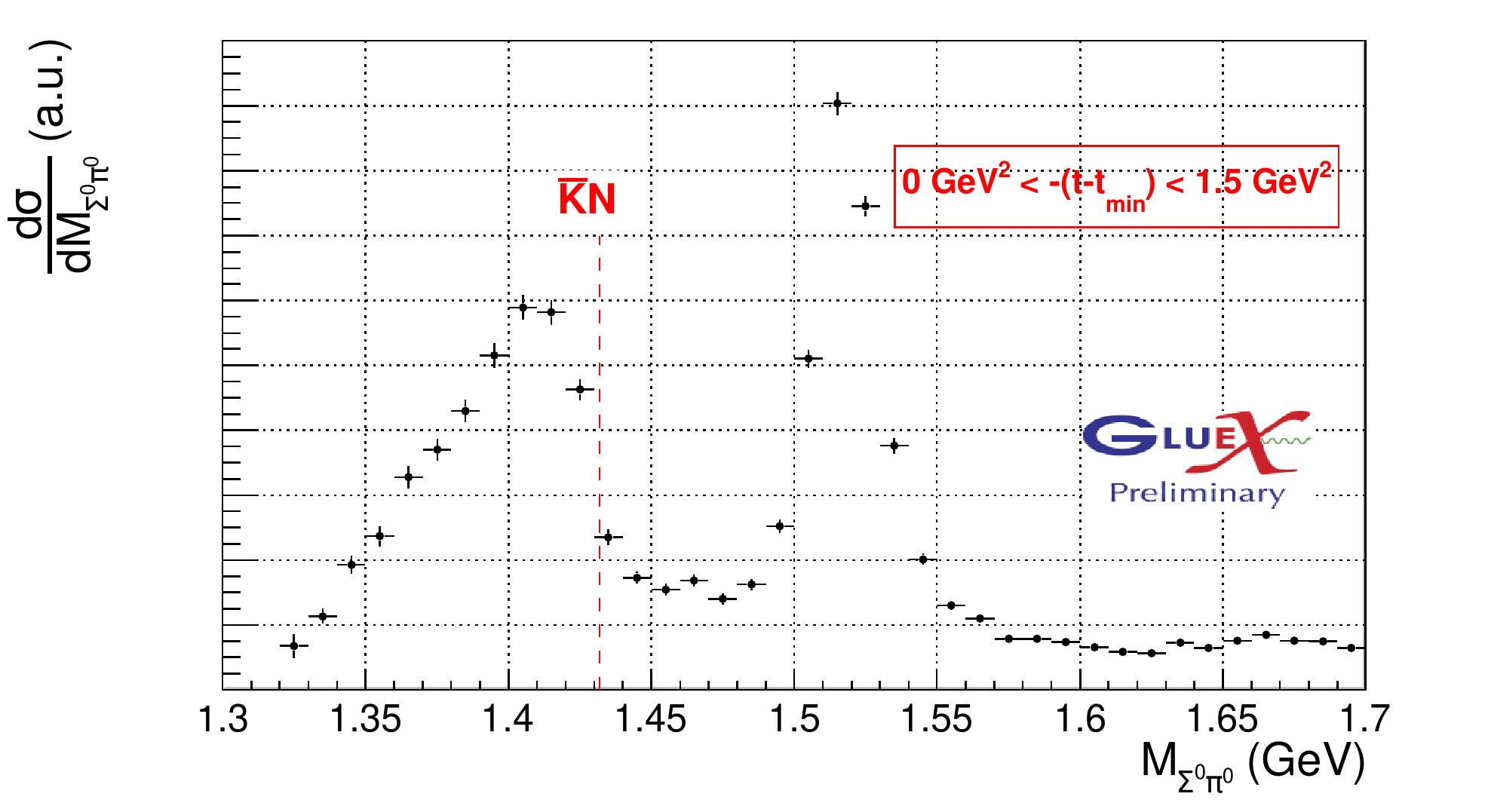}
  \captionsetup{justification=centering}
  \caption{}
  \label{fig:dsigma_dm}
\end{subfigure}

\caption{Invariant mass spectrum of $\Sigma^{0}\pi^{0}$ (a) for selected $\gamma p \to K^+\Sigma^0\pi^0$ events and (b) the differential cross section $d\sigma/dM_{\Sigma^0\pi^0}$ in the lower-mass region shown without a numerical scale since the absolute normalization has not been finalized. Red vertical dashed lines indicate the $N\bar{K}$ threshold.}
\label{fig:IM_dsigma}
\end{figure}

Figure~\ref{fig:IM_dsigma}(a) shows the invariant mass of $\Sigma^{0}\pi^{0}$ after applying all event selection criteria. Clear peaks of $\Lambda(1405)$ and $\Lambda(1520)$ are seen. There are $13351\pm139$ counts in the $\Lambda(1405)$ mass region ($M_{\Sigma^0\pi^0} < 1.47$~\text{GeV}). A sharp drop of yield for $\Lambda(1405)$ is seen at the $N\bar{K}$ threshold. This indicates the $\Sigma^0\pi^0$ mass resolution is good to resolve such threshold effects. From detector simulation it was found that the $\Sigma^0\pi^0$ mass resolution is $\sim$5 MeV in the $\Lambda(1405)$ mass region.

\subsection{Results}
\label{Sec:results}

\begin{figure}[h]
\centering
\includegraphics[width=12cm,clip]{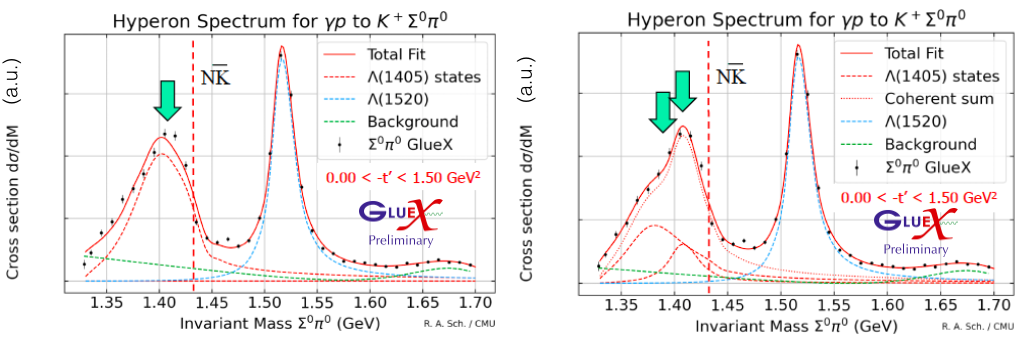}
\caption{Comparison of two hypotheses for the $\Lambda(1405)$ line shape, differing by having either one or two coherent Flatt\'{e} amplitudes in the fits. Green arrows indicate the centroids.  }
\label{fig:L1405_fit_compare}      
\end{figure}

In order to extract differential cross section $d\sigma/dM_{\Sigma^0\pi^0}$, the acceptance in the $\Sigma^0\pi^0$ mass range from threshold up to 1.7 GeV was found using simulated events which were analyzed in the same way as data. Figure~\ref{fig:IM_dsigma}(b) shows the  distribution for $d\sigma/dM_{\Sigma^0\pi^0}$ for $ 0 < -(t-t_{min}) < 1.5$~\text{GeV}$^2$. The uncertainties shown are statistical.
The absolute normalization of this measurement has not been finalized so the differential cross section is given without a numerical scale on the vertical axis. Therefore we report here on the line shape of the mass spectrum but not the absolute scale of the cross section. It can be seen  that the $\Lambda(1405)$ line shape appears to deviate from a simple Breit-Wigner form while the $\Lambda(1520)$ could be fit with a Breit-Wigner.
To test this hypothesis, we consider two fits to the $\Lambda(1405)$ line shape considering the $\Lambda(1405)$ to be described by a single Flatt\'{e} amplitude or by two such amplitudes (compound coherent $\Lambda(1405)$'s) as shown in Fig.~\ref{fig:L1405_fit_compare}. Table~\ref{tab-1} shows a summary of the different fitting procedures and results. The background parameterization takes into account the $Y^*(1670)$ contribution in addition to the linear background under $\Lambda(1405)$ and $\Lambda(1520)$ peaks. The two Flatt\'{e} amplitudes used in the $\Lambda(1405)$ two-state hypothesis were assumed to have a common Flatt\'{e} factor. Based on the $\chi^{2}$/\text{d.o.f.} of the fits the two-state hypothesis for $\Lambda(1405)$ is preferred to the one-state hypothesis. The centroids of two states are around $\approx$1387~\text{MeV and}\,$\approx$1409\,\text{MeV}.

\begin{table}[h]
\centering
\caption{Summary of two fit hypotheses used to fit $d\sigma/dM_{\Sigma^0\pi^0}$}
\label{tab-1}       
\begin{tabular}{ll}
\hline
One $\Lambda(1405)$ state & Two $\Lambda(1405)$ states \\\hline
Parameterized with one Flatt\'{e} amplitude & Two Flatt\'{e} amplitudes \\
Incoherent $\Lambda(1520)$ & Incoherent $\Lambda(1520)$  \\
Incoherent background & Incoherent background \\
$\chi^{2}$/\text{d.o.f.}\,\,\text{of the fit} = 5.1 & $\chi^{2}$/\text{d.o.f.}\,\,\text{of the fit} = 3.5 \\
\text{Centroid at}\,\,1407~\text{MeV} & \text{Two centroids at $\approx$1387~\text{MeV and}\,$\approx$1409\,\text{MeV}} \\\hline
\end{tabular}
\end{table}

In this simple model approach it is expected that unitarity is not preserved.  A more proper K-matrix approach to the problem is under development.   It enforces unitarity and analyticity of the resulting T-matrix and will result in proper identification of the T-matrix poles of the compound-state  $\Lambda(1405)$ region.

\begin{figure}[h]
\centering
\includegraphics[width=12cm,clip]{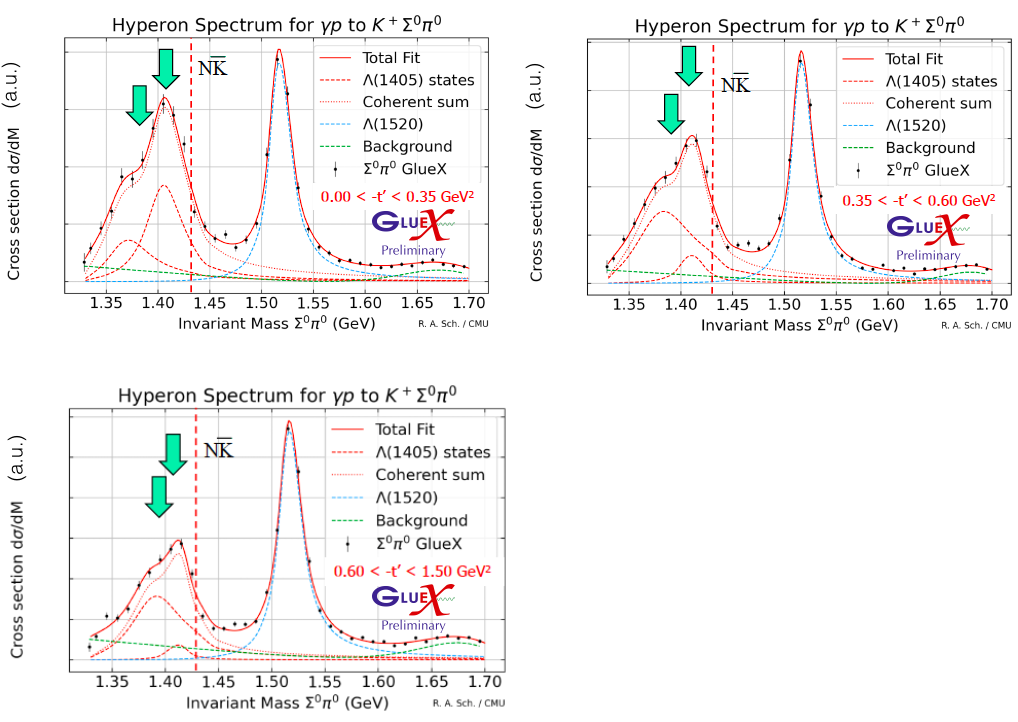}

\caption{The differential cross section $d\sigma/dM_{\Sigma^0\pi^0}$ in three bins of $-(t-t_{min})$ shown without a numerical scale since the absolute normalization has not been finalized. Red vertical dashed lines indicate the $N\bar{K}$ threshold. Green arrows indicate centroid positions.}
\label{fig:dsigma_dm_t_bins}      
\end{figure}

Differential cross sections $d\sigma/dM_{\Sigma^0\pi^0}$ were found for 3 different $-(t-t_{min})$ bins 0-0.35, 0.35-0.6 and 0.6-1.5 {GeV}$^2$. Figure~\ref{fig:dsigma_dm_t_bins} shows these distributions fit with the two-$\Lambda(1405)$ states hypothesis. Each of the fits seem to describe the data well and reproduce the drop of cross section at $N\bar{K}$ threshold. An notable feature is that the $\Lambda(1405)$ line shape seems to depend on $-(t-t_{min})$. The relative intensities of two $\Lambda(1405)$ states also seem to change with $-(t-t_{min})$. This possibly means that the contributions of the $N\bar{K}$ and $\Sigma\pi$ poles to $\Lambda(1405)$ production depend on $t$.  
Since the $\Lambda(1405)$ couples to both $\Sigma\pi$ and $N\bar{K}$ channels, a full description of the state should include both of these final states. GlueX has data for both $\gamma p \to K^+\Sigma^0\pi^0$ and $\gamma p \to K^+K^-p$ channels. A coupled-channel fit approach is in progress with using additional measurements from $K^-p$ data.

\newpage
\section{Conclusion}
\label{Sec:conclusion}

The GlueX detector is well suited for exclusive reconstruction of $\Lambda(1405) \to \Sigma^0\pi^0$. GlueX-I data provides the highest statistics obtained for this decay mode with $\sim$5 MeV resolution for $\Sigma^0\pi^0$ mass in the $\Lambda(1405)$ region. The extracted $\Lambda(1405)$ line shape clearly shows deviation from a single Flatt\'{e}-type Breit-Wigner amplitude.  Our preliminary fits to the line shape favor a picture with two coherent states for the $\Lambda(1405)$ region, and support previous theoretical and experimental evidence suggesting the $\Lambda(1405)$ is a composite baryon state.  The differential cross sections $d\sigma/dM_{\Sigma^0\pi^0}$ indicates a possible $t$-dependence to the $\Lambda(1405)$ line shape.  More elaborate fits using a K-matrix ansatz and data from $N\bar{K}$ final states are ongoing.  

\section{Acknowledgements}
\label{Sec:Acknlg}

We would like to acknowledge the outstanding efforts of the staff of the Accelerator and the Physics Divisions at Jefferson Lab that made the experiment possible. This work was supported in part by the U.S. Department of Energy, the U.S. National Science Foundation, the German Research Foundation, Forschungszentrum Jülich GmbH, GSI Helmholtzzentrum für Schwerionenforschung GmbH, the Natural Sciences and Engineering Research Council of Canada, the Russian Foundation for Basic Research, the UK Science and Technology Facilities Council, the Chilean Comisión Nacional de Investigación Científica y Tecnológica, the National Natural Science Foundation of China and the China Scholarship Council. This material is based upon work supported by the U.S. Department of Energy, Office of Science, Office of Nuclear Physics under contract DE-AC05-06OR23177.

\end{document}